\documentclass
[aps,prd,twocolumn,superscriptaddress,nofootinbib]{revtex4-1}%
\usepackage{amsmath}
\usepackage{amsfonts}
\usepackage{amssymb}
\usepackage{graphicx}
\usepackage{latexsym}
\usepackage{hyperref}
\usepackage{color,epsfig}
\usepackage{ifpdf}
\usepackage{bm}
\usepackage{wasysym}
\usepackage[english]{babel}
\usepackage{braket}
\usepackage{slashed}
\usepackage{longtable,array,cancel}

\definecolor{nicered}{rgb}{0.7,0.1,0.1}
\definecolor{nicegreen}{rgb}{0.1,0.5,0.1}
\hypersetup{colorlinks,citecolor= nicegreen,linkcolor= nicered}

\newcommand{\Red}[1]{{#1}}

\newcommand{\be}{\begin{equation}}
\newcommand{\ee}{\end{equation}}
\newcommand{\bea}{\begin{eqnarray}}
\newcommand{\eea}{\end{eqnarray}}
\newcommand{\afb}{A^{t\bar t}_{FB}}
\newcommand{\afbh}{A_{h}}
\newcommand{\stt}{\sigma^{t\bar t}}
\newcommand{\stth}{\sigma_h}
\newcommand{\lsim}{\stackrel{<}{_\sim}}

\begin{document}

\title{Forward-backward  $t\bar t$ asymmetry from anomalous stop pair production}

\author{Gino Isidori}
\email[Electronic address:]{gino.isidori@lnf.infn.it} 
\affiliation{INFN, Laboratori Nazionali di Frascati, Via E. Fermi 40 I-00044 Frascati, Italy.}

\author{Jernej F. Kamenik}
\email[Electronic address:]{jernej.kamenik@ijs.si} 
\affiliation{J. Stefan Institute, Jamova 39, P. O. Box 3000, 1001  Ljubljana, Slovenia}
\affiliation{Department of Physics,
  University of Ljubljana, Jadranska 19, 1000 Ljubljana, Slovenia}

\date{\today}

\begin{abstract}
We analyse a simple Standard Model (SM) extension with only two new light fields:
a scalar partner of the top $\tilde t$ (with mass above $m_t$) 
and a light neutral fermion $\chi^0$ (with mass of a few GeV), 
coupled to SM quarks via a Yukawa interaction. 
We show that such model can lead 
to  a significant enhancement of the forward-backward
asymmetry in  $t\bar t$ production at Tevatron
via the additional $t\bar t$ pairs produced from 
$\tilde t \tilde t^\dagger$ decays.
The model satisfies existing constraints on 
new-physics searches both at low and high energies, 
and could even address the cosmological dark-matter abundance.
The implications for future searches at the LHC are briefly 
outlined. 
\end{abstract}

\maketitle

\section{Introduction}

There are recent experimental indications of an anomalously large forward-backward 
asymmetry (FBA) in top-antitop pair production at the Tevatron. 
The asymmetry is defined as
\be
\afb = \frac{\sigma_{\Delta y>0} - \sigma_{\Delta y<0}}{ \sigma_{\Delta y>0} + \sigma_{\Delta y<0}}\,,
\ee
where $\Delta y = y_t - y_{\bar t}$ is the difference in rapidity, 
and positive rapidity is measured in the direction of the colliding $p$. 
The most significant measurement of $\afb$ is due to the CDF collaboration 
which after subtracting backgrounds and performing the unfolding procedure to recover the asymmetry at the partonic level reports~\cite{Aaltonen:2011kc}
\be
\afb = 0.158 \pm 0.074\,,
\label{eq:afbtot}
\ee
to be compared with the NLO QCD prediction $(\afb)_{SM} = 0.058 \pm 0.009$~\cite{Almeida:2008ug}. 
The observed FBA is even larger at high invariant masses of the $t\bar t$ system, 
in particular 
\be
\afbh = \afb{}_{(m_{t\bar t}>450{\rm GeV})} = 0.475 \pm 0.114\,,
\ee
while QCD predicts $0.088 \pm 0.013$~\cite{Almeida:2008ug}.

At the same time,  the total inclusive top pair production cross-section is currently measured 
to be~\cite{Schwanenberger:2010un}
\be
\stt  =( 7.50 \pm 0.48 )~{\rm pb}~,
\ee
that is consistent with the most recent theoretical SM predictions
for this observable: $(7.2\pm 0.4)~\rm pb$~\cite{Kidonakis:2010dk} and 
$(6.4 \pm 0.4)~ \rm pb$~\cite{Ahrens:2010zv}. 
Another important experimental constraint is the $m_{t\bar t}$ distribution of the production cross-section. 
As pointed out in~\cite{Blum:2011up}, the most significant information at high $m_{t\bar t}$
is the one derived from the next-to-highest measured bin~\cite{Aaltonen:2009iz}
\be
\stth = \stt_{(700~{\rm GeV}<m_{t\bar t}<800~{\rm GeV})} = (80\pm37)~\rm fb\,,
\label{eq:sigmatth}
\ee
to be compared with $(\stth)_{SM}=80 \pm 8$~fb ~\cite{Ahrens:2010zv}.

Several authors have analysed the possibility that the large FBA reported by CDF 
is obtained by an anomalous $t\bar t$ production mechanism interfering with the corresponding 
SM process~\cite{Severals,Blum:2011up}. 
The interference of the two amplitudes maximizes the possible impact on 
$\afb$ while minimizing deviations from the SM in $\stt$.
However, we observe that the present cross-section measurements still leave some 
room for an additional (incoherent) production of $t\bar t$ pairs.
This motivates us to analyse a different mechanism to enhance $\afb$, namely an 
anomalous production of $t\bar t$~+~invisible particles, which 
do not interfere with any SM process.  

At the level of the inclusive 
measurement, the non-standard production (passing the experimental selection cuts) 
can still contribute of up to $13\%$ compared to the SM (using the higher 
SM prediction in~\cite{Kidonakis:2010dk} as a conservative reference value). 
The percentage of the new contribution could even rise at high $m_{t\bar t}$,
provided it does not exceed $50\%$ of the SM 
cross-section in $\stth$.
Ideally, if the anomalous production would carry a $100\%$ FBA, it could 
perfectly accommodate the measured value of the total FBA without 
violating the bound on the cross section.

The production of the $t\bar t$~+~invisible final state can be obtained by the 
pair production of ``top partners" --particles decaying into a top quark  and light invisible 
states-- that are naturally expected in several SM extensions. 
If the mass difference between the top partner and the top is sufficiently 
small, the missing energy carried by the invisible states is small and
the  $t\bar t$ pairs thus produced would pass the experimental cuts applied to 
identify  $t\bar t$ pairs in the SM.  A prototype of 
such scenario would be the fourth generation up quark or a vector-like fermionic top partner, 
decaying into a top and one or several light invisible particles. However, colored 
fermions have large QCD cross-sections with a vanishing FBA contribution at leading order: 
this makes them them unattractive candidates for our purpose.  
On the contrary, a scalar top partner ($\tilde t$) of mass around $200$~GeV decaying 
into a top and a single invisible particle ($\chi^0$) is still perfectly allowed~\cite{tinv}.
This is because the QCD production cross-section for scalar particles
proceeds mainly via $p$-wave and thus vanishes at threshold~\cite{Beenakker:2010nq}. 

The stop and a light neutralino are naturally present in supersymmetric extensions of 
the SM, making this scenario particularly attractive. As we will discuss 
below, in the minimal supersymmetric extension of the SM (MSSM) the FBA generated by 
this mechanism is vanishingly small, but it could become sizable in more
general frameworks. Without assuming a specific model, 
here we adopt a phenomenological bottom-up approach: 
we assume $\tilde t$ and $\chi^0$
to be the only light relevant non-standard particles,  and we determine the nature of 
their interactions mainly by looking at phenomenological constraints. 
As a remnant of $R$-parity in the MSSM, we also assume a discrete
$Z_2$ symmetry under which only these non-standard states are charged, 
such that they can only be produced in pairs and such that $\chi^0$ is 
a stable particle.

\section{Identification of the model}

In order to generate a large $\afb$, the differential partonic 
cross-section $u(p_1) \bar u(p_2) \to \tilde t(p_1^\prime) {\tilde t^\dagger}(p_2^\prime)$ 
should exhibit a large 
$\hat t$-odd dependence,\footnote{The $u\bar u$ initial partonic state is chosen because it largely dominates over $d\bar d$ at the Tevatron, especially at high invariant masses, 
while the $gg$ initial state cannot produce a FBA.} 
where $\hat t = (p_1^\prime -p_1)^2$. This dependence is not generated 
by the leading QCD contribution to $u\bar u \to \tilde t {\tilde t^\dagger}$.

A first approach to generate a sizable $\hat t$-odd dependence in 
$u\bar u \to \tilde t {\tilde t^\dagger}$ is via heavy mediators
(in either $\hat s$ or $\hat t$ channel). Integrating out the heavy mediators 
leads to an appropriate set of higher-dimensional 
effective operators coupling the up quarks to the $ t {\tilde t^\dagger}$ pair. 
Up to canonical dimension six, the relevant operators are
\be
\bar u u  \tilde t^\dagger \tilde t\,,~~~\bar u \gamma_5 u \tilde t^\dagger \tilde t\,,~~~\bar u \gamma_\mu u  \tilde t^\dagger \partial^\mu \tilde t\,,~~~\bar u \gamma_\mu \gamma_5 u  \tilde t^\dagger \partial^\mu \tilde t\,.
\ee
In this case the required $\hat t$-odd dependence can only appear in the numerator 
of the cross section, but the kinematical structure of all the above operators 
is such that they do not generate it. A non-vanishing $\hat t$ dependence can
appear only introducing operators of dimension 7 or higher, which are expected 
to be more suppressed by naive dimensional analysis.
Since the contributions with a $\hat t$-odd dependence should dominate 
the cross section, which receives also a sizable contribution from ordinary 
QCD, this approach appears to be highly contrived. 
 
A second approach is to assume light $\hat t$-channel
mediators for $u\bar u \to \tilde t \tilde t^\dagger$. In this way the $\hat t$-odd dependence 
is naturally induced by the $\hat t$-channel propagators 
 and gets more pronounced the lighter the mediators. The lightness of the mediators is also required to generate a sizable cross-section (larger than the QCD induced one) while 
maintaining perturbativity of the associated mediator couplings to (light) quarks 
and $\tilde t$. The mediators need to be neutral and, more generally, 
$SU(2)_L \times U(1)_Y$ singlets,
to avoid the bounds on new light states from LEP. Assuming they couple trilinearly to $u$ and $\tilde t$, they should also be fermions. Since we have already introduced a light neutral fermion in order to account for $\tilde t \to t \chi^0$, the simplest possibility 
is to assume $\chi^0$ itself to be the mediator. We are thus led to consider 
the following simple Lagrangian 
\bea
\mathcal L &=& \mathcal L_{SM} + (D_\mu \tilde t)^\dagger (D^\mu \tilde t) - m_{\tilde t}^2 \tilde t^\dagger \tilde t +  \bar \chi^0 (i \gamma_\mu D^\mu) \chi^0
 \nonumber\\
&&- m_\chi {\bar\chi}_c^{0}  \chi^0~
+ \sum_{q=u,c,t} (\tilde Y_{q} \bar q_R \tilde t \chi^0 + {\rm h.c.} )~,
\label{eq:Lagr}
\eea
where we have introduced effective $q_R\tilde t\chi^0$ couplings for  all three 
generations of right-handed up quarks. Having assumed  $\chi^0$ to 
be an $SU(2)_L\times U(1)_Y$ singlet, this implies that $\tilde t$
has the same $SU(2)_L\times U(1)_Y$ quantum numbers of right-handed up quarks.
As we will discuss below, beside the minimality of the new states introduced,
the choice of a pure right-handed coupling minimizes the impact 
on electroweak observables and flavour-changing neutral-current (FCNC) transitions.

Following the principle of introducing the minimal set of relevant new states, 
we have assumed $\chi^0$ to have only a Majorana mass term. This hypothesis 
has essentially no impact on collider phenomenology (provided $m_\chi$
is sufficiently small \Red{in order to suppress the production of same-sign tops}), 
while it may be an important ingredient if we require $\chi^0$ to be a 
viable dark-matter candidate (see below).

With this choice $\tilde t$ can be identified with the right-handed stop of the 
MSSM or, more generally,
with any combination of right-handed up-type squarks.
In principle, $\chi^0$ could be identified with the bino of the MSSM.  
However,  bino couplings are completely determined by electroweak gauge symmetries 
and ultimately turn out to be too small to significantly affect the FBA. 
On the other hand, \Red{we see no obstacles to accomodate 
the required framework in extensions of the MSSM with extra 
$SU(2)_L\times U(1)_Y$ singlets. 
}
An important condition in order to implement this mechanism 
in supersymmetric extensions of the SM is a sufficeintly heavy gluino mass, 
such that gluino-mediated amplitudes do not provide a significant enhancement of the 
$t \tilde t^\dagger$ production cross section.

As anticipated, we keep our discussion general analysing the phenomenological
implications of (\ref{eq:Lagr}) without assuming extra model-dependent 
conditions. In order to avoid the production of $\chi^0$ pairs in charm 
decays we require $m_\chi > m_D/2$, while for the $\tilde t$ state 
we require $m_{\tilde t} > m_t$ and we impose the contraints derived in~\cite{tinv} for top-partner pair production at Tevatron.

\section{Collider phenomenology}

As an illustration of the resulting phenomenology at the Tevatron, 
we choose $m_{\tilde t} = 200$~GeV and $m_{\chi^0}=2$~GeV. We simulate our signal $p\bar p \to \tilde t \tilde t^\dagger \to t\bar t \chi^0\chi^0$ at the partonic level with MadGraph/MadEvent 4.4.57~\cite{Alwall:2007st} and using CTEQ6L1 set of PDFs~\cite{Pumplin:2002vw}. 
For $\tilde Y_{u}\simeq 0.85$ we find that the total $\tilde t \tilde t^\dagger$ 
production cross-section reaches $12\%$ of the SM  
$t\bar t$ cross-section, exhibiting a $50\%$ FBA. 
Assuming $\tilde t$  decays dominantly into $t\chi^0$ (or assuming $\tilde Y_t \gg \tilde Y_{u,c}$)
and applying the  $t\bar t$ reconstruction cuts, 
this leads to a predicted $\afb$ 
within one standard deviation from the experimental data in (\ref{eq:afbtot}).
At the same time, with this parameter choice  the value of the total cross-section 
is within one standard deviation from the measured one, using the 
prediction in~\cite{Kidonakis:2010dk} as a conservative SM normalization.
Some tension does develop only when comparing the high invariant-mass data, where 
the predicted cross-section and FBA tend to be higher and smaller than data, respectively.
In particular, the predicted $\stth$ exhibits a tension of $1.6~\sigma$ when compared with 
(\ref{eq:sigmatth}), while he SM tension of $3.4~\sigma$ in $\afbh$ is 
only relieved to around $1.9~\sigma$. 

\begin{figure}[t]
\includegraphics[scale= 1.3]{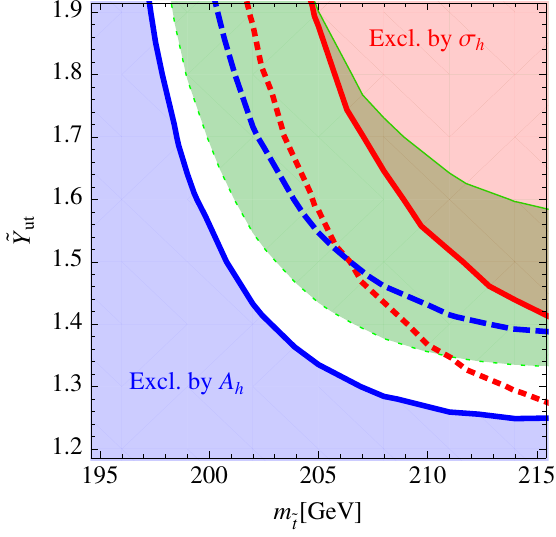}~~~~~~~~
\caption{Tevatron constraints in the $m_{\tilde t}$--$\tilde Y_u$ plane.
The inclusive $\afb$ and $\stt$ are reproduced within $1~\sigma$ in the 
central green band. The region below the continuous (dashed) blue 
line is excluded by $\afbh$ at 95\% C.L. (90\% C.L.). The region above the continuous (dotted) 
red line is excluded by $\stth$ at 95\% C.L.~(90\% C.L.).}
\label{fig:parSpace}
\end{figure}

In general, as long as $\chi^0$ is much lighter than $\tilde t$ and $\tilde Y_t \gg \tilde Y_{u,c}$, the results do not depend on the precise values of $m_{\chi^0}$ and $\tilde Y_{t}$. For this reason, we have performed a more accurate evaluation 
of the various constraints on the parameter space of the model at fixed $m_{\chi^0}$ and $\tilde Y_{t}$.
In particular, we have set $m_{\chi^0}=2$~GeV and $\tilde Y_{t}=4$ (the largest value to ensure a perturbative behavior 
in $\tilde Y_{t}$) and varied the remaining relevant parameters, namely $\tilde Y_u$ and $m_{\tilde t}$. 
Since the condition $\tilde Y_t \gg \tilde Y_{u}$ is not always fulfilled, we have taken into account  
the actual $\tilde t\to t\chi^0$ branching ratio when computing our signals. 
The resulting comparison with the relevant constraints coming 
from $t\bar t$ production at the Tevatron is displayed 
in Fig.~\ref{fig:parSpace}.

As shown in Fig.~\ref{fig:parSpace}, the requirements of a large 
FBA and a small impact on the cross section are partly in 
conflict, especially at high invariant masses.
However, there is a region of the parameter space 
where all constraints are satisfied at the  $90\%$ C.L.
This happens for $200\lesssim m_{\tilde t}\lesssim 205$~GeV 
and $\tilde Y_{u}\gtrsim 1.5$ (part of the green area 
where the dotted line is above the dashed one). 
The reason for that lies in the interplay of the $\tilde Y_{u}$ 
dependence in the $\chi^0$-mediated cross-section generating 
a large FBA and in $Br(\tilde t\to t\chi)$. In particular, 
for $|\tilde Y_u|\lesssim |\tilde Y_t|$ the 
$\hat t$-even QCD contribution is suppressed by $Br(\tilde t\to t\chi)$
more than the $\hat t$-odd contribution, 
which enables us to obtain a larger FBA value for a given contribution to the cross-section. 

\Red{
The two additional invisible particles in the final state result in extra missing energy compared to the SM $t\bar t$ production.  This might result in (1) observable modifications of the various kinematical distributions used in the experimental analyses of the FBA (as well as the $t\bar t$ production cross-section and the top quark mass) to discriminate signal from background; (2) tensions between the leptonic, semileptonic and all-hadronic modes in the precise measurements of the $t\bar t$ cross-section and the top quark mass. 
Unfortunately, most present top quark analyses employ multivariate techniques which are difficult to reproduce without the full detector simulation. Nonetheless, we have estimated the size of such effects by comparing the leptons and jets $p_T$ distributions as well as distributions of missing transverse energy ($E_T^{\rm miss}$) and the scalar sum of transverse particle energies ($H_T = \sum_i E^{i}_T$) in the process $p\bar p \to t\bar t \chi^0 \chi^0$ (at preferred values of the model parameters) with those of the SM $t\bar t$ production. We have considered these distributions separately for the leptonic mode with kinematical cuts used in~\cite{CDF10163}, the semileptonic reconstruction channel described in~\cite{Strycker} as well as the all-hadronic mode used recently for the top mass measurement in~\cite{CDF10456}. We have simulated both signals using MadGraph, interfaced with Pythia 6.4.14~\cite{Alwall:2008qv} for showering and hadronization, and the CDF PGS for detector simulation with a cone ($dR=0.4$) jet reconstruction algorithm.  In the leptonic mode, we first observe that the fraction of the $t\bar t \chi^0 \chi^0$ signal passing the kinematical cuts is 
comparable to the SM $t\bar t$ signal (the same is true also for the semileptonic mode). Regarding the kinematical distributions outlined above, the effects of the extra $\chi^0$'s can be readily described as a shift in the distributions by few GeV. Since such shifts only affect the $O(10\%)$ new contribution to the total $t\bar t$ sample, the resulting effects turn out to be much smaller than the reported systematic and statistical uncertainties in these distributions. As a consequence, 
in di-leptonic semple  we expect the modifications 
in the total measured $t\bar t$ cross-section or the top-quark mass to be  
smaller than the present statistical uncertainties.  
In the semileptonic case, the shift in the distributions is a bit more pronounced,
of the order of $10\,\rm GeV$,  but the resulting change in the 
combined $t\bar t \chi^0 \chi^0 + t\bar t$ signal sample distributions is still smaller than the associated statistical uncertainties. We have also compared the reconstructed transverse leptonic $W$ masses ($m_T^{W} = \sqrt{2 E_T^{\rm miss} E_T^{\ell} [1-\cos(\phi_\ell - \phi_{E_T^{\rm miss}})]}$, where $\phi_i$ are the corresponding azimuthal angles) and found good agreement between the distributions of the two signals. 
Most important, the distribution peak in the $t\bar t \chi^0 \chi^0$ sample does not appear shifted with respect to the SM $t\bar t$ case. Therefore, 
we expect that resulting effects on $t\bar t$ cross-section or top-quark mass measurements 
using the semileptonic mode to be within the present statistical and systematic uncertainties. 
Finally, for the all-hadronic mode, we find that the requirement of no significant $E_T^{\rm miss}$ reduces the fraction of $t\bar t \chi^0 \chi^0$ signal events passing the cuts to one third, 
compared to the SM $t\bar t$ sample (assuming an  uncertainty in $E_T^{\rm miss}$ reconstruction of $\sigma_{E_T^{\rm miss}}\simeq 0.5 \sqrt{H_T}\, {\rm GeV}^{1/2}$~\cite{CDFET}). The resulting discrepancy 
in the measured total $t\bar t$ production cross-section compared to the semileptonic and leptonic 
modes for preferred values of our model parameters would be of the order of $7\%$, 
that is less than the present uncertainties of the individual analyses. 
}

\medskip 

Recently top-pair production production has also been measured at the LHC. 
The value reported in~\cite{Khachatryan:2010ez} for the total $t\bar t$ cross section 
in $pp$ collisions at $\sqrt s = 7$~TeV center-of-mass energy is 
\be
\sigma^{t\bar t}_{LHC} = (194 \pm 78)~\rm pb\,,
\ee
in agreement with the SM prediction of $158\pm 24$~pb~\cite{Campbell:2010ff}. 
For our preferred choice of parameters the corresponding $\tilde t\tilde t^\dagger$ production cross-section 
is only around $10$~pb~\cite{Beenakker:2010nq}\footnote{At the LHC, the QCD production through $gg$ initial state completely dominates over the anomalous $\hat t$-channel contributions for our choices of parameters.}, 
still below the present sensitivity of the LHC experiments.

The $\tilde t \tilde t^\dagger$ pairs decaying to two up quarks 
and two $\chi^0$'s contribute to the 
jets plus missing transverse energy ($E^{\rm miss}_T$) signatures, searched for both at the Tevatron~\cite{Aaltonen:2008rv} and at the LHC~\cite{Khachatryan:2011tk,ATLASSUSY}.  At present the most sensitive search is  
the ATLAS analysis of two jets plus $E^{\rm miss}_T$~\cite{ATLASSUSY}, 
based on $35\,\mathrm{pb}^{-1}$ of data, and in particular 
the scenario $A$ of~\cite{ATLASSUSY}, which is 
optimized for low-mass squark pair production. 
Here two jets with $p_T>140,~40$~GeV respectively, 
and $E^{\rm miss}_T>100$~GeV
are required. Additional kinematic cuts are also imposed,
among which the most relevant are 
$m_{eff}=\sum_{i=1}^2 |p^i_T| + E^{\rm miss}_T > 500$~GeV 
and $E_T^{\rm miss}/m_{eff}>0.3$. Under these conditions 
$87$ events are found passing all the cuts, 
with a SM background estimation of $118\pm 42$. 
Again we have simulated our signal contribution using MadGraph, interfaced with Pythia 6.4.14~\cite{Alwall:2008qv} for showering and hadronization and the ATLAS PGS for detector simulation with a $k_T$ jet reconstruction algorithm. 
For the most interesting region of the model's parameter space, 
we find that 
the signal cross-section passing the $p_T$ and $E_T^{\rm miss}$ cuts alone
is at the level of $1$~pb or less, still below the present experimental sensitivity.

\medskip

In principle, also the Tevatron 
analyses for single top production 
could be used to set bounds 
on the parameter space of our model. However, 
in this case it is very difficult to assess the sensitivity. 
Present experimental analyses employ sophisticated multivariate discriminants including tight constraints on the missing transverse energy distribution, which is required to agree with expectations from a single, leptonically decaying $W$ boson in the final state~\cite{Aaltonen:2010jr}. Here we simply note that  
the $\tilde t \tilde t^\dagger$ production is small compared to the
main SM backgrounds, namely $W$+jets and $t\bar t$ production, 
which have a similar signature. Thus we do not expect very stringent
constraints from the present single-top production searches. 

\Red{If $\chi$ is a Majorana particle, then in priciple one can expect 
a non-vanishing production of same-sign tops from 
the $uu \to \tilde t \tilde t$ process; however, this amplitue
is strongly suppressed by the smallness of $m_\chi$. For 
$m_\chi = 2$~GeV and $m_{\tilde t} > m_t$ the production of 
same sign tops is suppressed by $(m_\chi/m_{\tilde t})^2 \sim 10^{-4}$
with respect to  the $\chi$-mediated
stop anti-stop production (that is a small fraction
of the SM $t\bar t$ cross section). Beside the $(m_\chi/m_{\tilde t})^2$ power 
suppression, the same-sign top cross section receives 
logarithimic enhencements both near threshold and 
at large dilepton invariant masses; however, 
these are compensated by the small probability distribution for 
the initial $u u$ state.  As a confirmation of this qualitative evaluation,
we have perfomed a quantitative study of the  same-sign top
cross section for our preferred range of parameters
 finding $\sigma(tt)_{\rm Tevatron} \sim 0.01$~fb
and $\sigma(tt)_{\rm [LHC\, 7\, TeV]} \sim 3$~fb: a level 
well below the present sensitivity.}

\section{Flavour physics and Dark matter}

A very significant constraint on $\tilde Y_q$ comes from the $D^0$--$\bar D^0$ mixing amplitude, 
to which $\chi^0$ and $\tilde t$ can contribute at one loop. 
Introducing the following low-energy effective dimension-six Hamiltonian 
\be
{\cal H}_{\rm eff} = C^R_{ud}~(\bar c_R \gamma_\mu  u_R)^2,
\ee
the one-loop induced contribution by our Lagrangian in (\ref{eq:Lagr})  is 
\be
C^R_{ud} = - \frac{1}{32\pi^2 m_{\tilde t}^2} (\tilde Y_{c} \tilde Y_{u}^*)^2~.
\ee
Using the bound $|C^R_{ud}| < 1.2 \times 10^{-3}$~TeV$^{-2}$~\cite{Isidori:2010kg}, 
this implies 
\be
\left|\tilde Y_{c}/\tilde Y_{u} \right| < 0.06~,
\ee
for our preferred values of $m_{\tilde t}$ and $\tilde Y_{u}$. 
This is certainly a fine-tuned condition, although it is 
comparable to the hierarchies exhibited by the 
Yukawa couplings within the SM.

Another interesting consequence of our scenario are FCNC top quark decays to a light quark jet and missing energy (i.e. $t \to u \chi^0 \bar \chi^0$). Neglecting the $\chi^0$ mass dependence, the decay rate is given by
\be
\Gamma(t \to u \chi^0 \bar \chi^0) = \frac{|\tilde Y_t \tilde Y_u|^2 m_t^5}{6144 \pi^3 m_{\tilde t}^4}\,.
\ee
For our illustrative choice of parameters, the branching ratio might reach 
the level of $10^{-3}$, which could be within the projected LHC sensitivity~\cite{Li:2011ja}.

\Red{Finally, it is interesting to note that stable fermions like $\chi^0$
with mass of a few GeV, annihilating to light quark pairs via effective dimension six operators 
have been previously considered as dark-matter candidates~\cite{Beltran:2008xg}. 
In our case, assuming for the moment that $\chi^0$ is a Dirac fermion, 
the low-energy coupling  of $\chi^0$ to light quarks induced by the $\tilde t$ exchange 
leads to 
\be
\mathcal L^{\rm eff}_{\rm annih.} = \frac{|\tilde Y_u|^2}{4 m_{\tilde t}^2} \bar u_R \gamma_\mu u_R \bar \chi^0 \gamma^\mu (1-\gamma_5) \chi^0\,.
\label{eq:DM}
\ee
In this case the dominant contribution to the thermal  annihilation rate  of $\chi^0$ comes 
from the vector current part ($\bar\chi^0 \gamma_\mu \chi^0$) of the above operator. 
Using the results of~\cite{Beltran:2008xg}, and 
setting $m_{\tilde t}\simeq 200$~GeV and $|\tilde Y_u|\simeq 1.5$
in (\ref{eq:DM}) we find that the correct relic abundance 
of $\chi^0$  is reproduced for $m_{\chi^0}\simeq 2$~GeV,
that would perfectly fit with the collider data. Such contributions are however in some tension 
with existing direct dark-matter detection experiments~\cite{Ahmed:2010wy}.   Furthermore, 
a recent dark-matter search using the
 WMAP CMB~\cite{Hutsi:2011vx} spectrum disfavors thermal relics of masses below $5$~GeV. 
Therefore in order for $\chi^0$ to have escaped these existing searches, its thermal annihilation 
cross-section has to be somewhat smaller than suggested by the dark-matter relic abundance measurements.  
This could happen if $\chi^0$ is of Majorana type, 
so that its thermal annihilation cross-section is velocity suppressed.
Alteratively, we could assume that $\chi^0$ represents only a fraction of the total 
relic abundance. A third possibility is to lower the $\chi^0$ mass below the sensitivity 
range of the present direct detection experiments, ignoring the cosmological bounds.
If $\chi^0$ is a Majorana particle, then only the axial current ($\bar\chi^0 \gamma_\mu \gamma_5 \chi^0$)
part of (\ref{eq:DM}) is allowed and current dark-matter searches (both direct and indirect)
do not exclude a particle of this type with mass of a couple GeV or less
 (see e.g.~\cite{Beltran:2008xg,Ahmed:2010wy,Goodman:2010yf}). 
In this case the correct dark-matter abundance can in principle still be obtained if $\chi^0$ is produced non-thermally 
as in the asymmetric dark-matter scenarios~\cite{Kaplan:2009ag}, for example through the out-of-equilibrium 
decays of heavier particles, not included in our low-energy model.}

\section{Conclusions} 

Our knowledge of top-quark physics is still rather limited. 
As we have shown, it could well be that the $t \bar t$ sample 
analysed at the Tevatron is enriched by non-standard contributions 
coming from decays of a scalar top-partner, 
with the electric and color charge of the top quark.
The top partners should be produced in pairs
($\tilde t \tilde t^\dagger$), have a mass slightly above $m_t$, 
and a large $\tilde t \to t\chi^0$ branching ratio, where 
$\chi^0$ is a SM singlet with mass of a few GeV or less
(escaping detection).
With a proper tuning of the $q\tilde t\chi^0$  effective couplings,
the additional  $t \bar t$  pairs produced in this way could 
account for the large $\afb$ observed at CDF. 

The simple model we have proposed, where  $\tilde t$ and 
$\chi^0$ are the only relevant new light states, is fully
consistent with present high-energy data 
and contains a dark-matter candidate.
\Red{The model requires a non-trivial flavour structure,\footnote{~A flavour-violating 
structure of this type has also been postulated in~\cite{Shelton:2011hq}, 
although in a different 
framework, and could be attributed to a specific flavour symmetry.} such that 
$\tilde t$ has large couplings to both $\chi^0t_R$ and $\chi^0u_R$,
and a vanishing coupling to $\chi^0c_R$.  
However, it is quite natural as far as the field content is concerned:
a light scalar partner for the top is a natural candidate for 
a stabilization of the leading quadratic divergence on the 
SM Higgs mass term. In this paper we have pursued a
bottom-up phenomenological approach and 
we have not analysed how this simple framework could be 
extended at high energies to be considered a viable extension 
of the SM; however, at this stage we see no obstacles 
to consider it as the low-energy side of a more ambitious 
supersymmetric model. }

Interestingly, this framework can soon be tested in more detail 
at the LHC. In particular: i) the large $E^{\rm miss}_T$ of the 
sub-leading $\tilde t \to u \chi^0$ decay mode, and 
ii) the rise of the $pp\to \tilde t \tilde t^\dagger \to t\bar t$ 
QCD cross-section at large $m_{t \tilde t^\dagger}$,
offer powerful tools to disproof or find 
evidences of this non-standard framework.

\section{Acknowledgments}  
We are grateful to Martti Raidal for clarifying an ambiguity in the dark-matter 
discussion in the first version of the paper.
G.I.~acknowledges the support of the Technische Universit\"at M\"unchen 
-- Institute for Advanced Study, funded by the German Excellence Initiative.
J.F.K.~thanks the INFN Laboratori Nazionali di Frascati where part of this work was completed for their hospitality and acknowledges financial support by the Slovenian Research Agency.

\section{Added note}
After this work was completed, we become aware of a new anomaly reported
by the CDF experiment, namely an excess in the production of jet pairs 
in association with a W boson~\cite{Aaltonen:2011mk}. In particular, 
CDF reports an excess in the di-jet invariant mass ($M_{jj}$) distribution 
for $120 \lsim M_{jj} \lsim 160$~GeV. 
We note that 
our model predict an excess over the SM prediction in the $W+jj$ channel
due to the sub-leading $\tilde t \to u +\chi$ decay mode:
$\tilde t \tilde t^\dagger \to u \bar t~(t \bar u) + E_{\rm miss} \to  u \bar b~(b \bar u) + W^- (W^+) + E_{\rm miss}$.
We have simulated this decay chain and found that the resulting 
$M_{jj}$ distribution is indeed peaked around $140-150$~GeV
for our preferred range of parameters. However, the size of the
cross-section is substantially smaller (about 1/3) with respect 
to the central value of the non-standard effect reported by CDF
(after taking into account the experimental cuts applied in~\cite{Aaltonen:2011mk}). 
Moreover, the $M_{jj}$ distribution predicted in our framework
is substantially broader, with  $\sigma(M_{jj})\sim 100$~GeV
with respect to the one reported in~\cite{Aaltonen:2011mk}.
From these considerations we conclude that the inclusion of 
the $\tilde t \tilde t^\dagger$ production could partially
improve the agreement with data also in the $W+jj$ channel,
although a quantitative evaluation of this improvement 
 would require a more accurate evaluation of 
detector efficiencies and resolutions, especially on the jet
variables, that is beyond the scope of the present work.

\end{document}